\documentstyle[psfig,floats,twocolumn,eqsecnum,prb,aps]{revtex}
\begin{document}
\draft

\twocolumn[\hsize\textwidth\columnwidth\hsize\csname
@twocolumnfalse\endcsname

\title{Microwave Conductivity due to Scattering from Extended Linear Defects \\
in $d$-Wave Superconductors}
\author{Adam C. Durst and Patrick A. Lee}
\address{Department of Physics, Massachusetts Institute of Technology,
Cambridge, Massachusetts 02139}
\date{\today}

\maketitle

\begin{abstract}
Recent microwave conductivity measurements of detwinned, high-purity,
slightly overdoped YBa$_{2}$Cu$_{3}$O$_{6.993}$ crystals reveal
a linear temperature dependence and a near-Drude lineshape for
temperatures between 1 and 20 K and frequencies ranging from 1 to 75 GHz.
Prior theoretical work has shown that simple models of scattering by
point defects (impurities) in $d$-wave superconductors are inconsistent with
these results.  It has therefore been suggested that scattering by extended
defects such as twin boundary remnants, left over from the detwinning
process, may also be important.  We calculate the self-energy and microwave
conductivity in the self-consistent Born approximation (including
vertex corrections) for a $d$-wave superconductor in the presence of
scattering from extended linear defects.  We find that in the
experimentally relevant limit ($\Omega, 1/\tau \ll T \ll \Delta_{0}$),
the resulting microwave conductivity has a linear temperature
dependence and a near-Drude frequency dependence that agrees well
with experiment.
\end{abstract}

\pacs{PACS numbers: 74.72.Bk, 72.10.Fk, 61.72.Mm, 78.70.Gq}
]

\section{Introduction}
\label{sec:intro}
Of the numerous phases of the high-$T_{c}$ cuprate phase
diagram, the low temperature superconducting phase
appears to be the least mysterious.  All indications point
to an order parameter of $d_{x^2-y^2}$ symmetry and well
defined quasiparticle excitations above the condensate.
\cite{lee97a,ore00}
An excellent probe of the low energy excitations of this
$d$-wave superconductor is the measurement of low temperature
microwave conductivity.  Experimentally, the real part of the
microwave conductivity, $\sigma(\Omega,T)$, can be
extracted from microwave measurements of complex surface
impedance $Z_s(\Omega,T)$.
\cite{hos99,bon96,lee96,bon93,bon92}
Theoretically, at low temperatures,
it is determined by calculating the linear response of a
$d$-wave superconductor in the presence of scattering due to
static disorder.
\cite{hir93,hir94,hir89,hir86,mon87,hir88,gra95,sun95}
(Low temperature is defined as low enough such that all inelastic
scattering \cite{wal00} is negligible.)  The mystery
is that, in this least mysterious of phases, the simplest
theoretical models do not agree with experiment. \cite{ber00}

The microwave conductivity of detwinned, high-purity, slightly
overdoped YBa$_{2}$Cu$_{3}$O$_{6.993}$ was measured by
Hosseini {\it et al.} \cite{hos99}  Data was taken at five
frequencies between 1 and 75 GHz for temperatures from 1 to
95 K.  Below 20~K, in the regime dominated by elastic scattering,
they find that the microwave conductivity has an approximately linear
temperature dependence and a near-Drude frequency dependence.
Fitting the data to a Drude form, they extract a spectral weight
that is linear in temperature (in agreement with the measured
superfluid density) and an effective scattering rate, $1/\tau$,
that is approximately constant over the entire low temperature
regime (1 to 20 K).

From the extracted value of the scattering rate
($5.6 \times 10^{10} \,\mbox{s}^{-1}$) and
the frequency and temperature range of the experiments,
it can be seen that these measurements correspond to the
parameter regime $\Omega,1/\tau \ll T \ll \Delta_{0}$
where $\Delta_{0}$ is the gap maximum.  For this parameter
regime, the theoretical picture is as follows.  A $d$-wave
superconductor is characterized by an order parameter that
vanishes linearly at each of four gap nodes.  For $T \ll \Delta_{0}$,
transport is dominated by low energy quasiparticle excitations
generated in the vicinity of the nodes.  For $1/\tau \ll T$,
the quasiparticles are generated thermally, rather than by the
presence of disorder. \cite{lee93}  (Hence, in what follows, we refer to
this as the {\it thermal} regime.)  The most straightforward model for
such a system is one in which thermally generated nodal quasiparticles
are scattered due to the presence of point defects such as impurities.
Performing an impurity average and using a Kubo formula approach
within the self-consistent $t$-matrix approximation, the self-energy
and electrical conductivity have previously been calculated.
\cite{hir93,hir94,hir89,ber00}

Surprisingly, the predictions of such calculations do not agree with
the experimental results.
The trouble comes in attempting to reproduce the constant scattering
rate extracted from experiment.  For a $d$-wave superconductor
in the thermal regime, the anisotropic Dirac excitation spectrum
yields a quasiparticle density of states that is linear in energy.
This strong energy dependence of the density of states is then
directly reflected in the results of the impurity $t$-matrix
calculations.  Such calculations yield $1/\tau \sim T$
and $\sigma \sim \mbox{const}$ in the Born limit and
$1/\tau \sim 1/T$ and $\sigma \sim T^{2}$ in the unitary limit.
More generally, over the full range of scattering strengths, it
has been shown that simple models \cite{hir00} of point defect
(impurity) scattering are inconsistent with experiment. \cite{ber00}

Despite this, we certainly expect the presence of impurities to make a
significant contribution to the total scattering.  However, the
observed disagreement between theory and experiment
suggests that an additional scattering mechanism
may also be important.  In what follows, we propose that this additional
mechanism is scattering from extended linear defects.
Our motivation for doing so is twofold.  First of all, we note
that scattering from a series of parallel lines differs from point
defect scattering in an important way.  While point defects scatter
in all directions, line defects only scatter in the direction normal
to their orientation.  Hence, whereas point defect scattering samples
the full quasiparticle density of states, line defect scattering restricts
final states to those for which there is no change in the parallel component
of momentum.  Therefore, the line defect scattering rate need not inherit
the strong energy dependence of the density of states and is thereby
capable of exhibiting the less energy dependent form required to agree
with experiment.  Secondly, we note that upon growth, YBCO crystals are full
of ``lines'', twin boundary lines.  Although the twin boundaries are
subsequently removed via a detwinning procedure, remnants of their
structure may be left behind in the form of extended linear defects.
Since lines are prevalent in this material and line defects tend to
exhibit features of the measured behavior, it is logical that we should
examine the contribution made by line defect scattering.
For simplicity, we consider herein a system where extended linear
defects provide the sole scattering mechanism, leaving the analysis of
a system with both line defects and point defects for future study.

In Sec.~\ref{sec:crystal}, we discuss twinning in YBCO and the nature
of the detwinning process.  In light of the observed twinning structure
and guided by the notion that the detwinning process leaves behind remnants
in the form of line defects, we develop, in Sec.~\ref{sec:extended},
a model for quasiparticle scattering from
extended linear defects in a $d$-wave superconductor.
In Secs.~\ref{sec:selfenergy} and \ref{sec:microcond}, we calculate
the resulting self-energy and microwave conductivity.
For parameter values within the thermal regime, we obtain analytical
results which are compared to experiment in Sec.~\ref{ssec:analytical}.
Numerical results, including deviations from our analytical expressions and
valid beyond the thermal regime, are presented in Sec.~\ref{ssec:numerical}.
Conclusions are discussed in Sec.~\ref{sec:conclusions} where we
provide physical motivation for our calculated results and discuss the
implications of our findings.

\section{Twin Boundaries in YBCO \cite{HAR01}}
\label{sec:crystal}
Twinning occurs naturally during the growth of oxygen deficient
YBa$_2$Cu$_3$O$_{7-\delta}$.  At full oxygenation ($\delta=0$), the
crystal structure is orthorhombic at room temperature with lattice
parameters $a=3.227~\mbox{\AA}$ and $b=3.8872~\mbox{\AA}$, the latter being the
orientation parallel to the CuO chain layer. \cite{sha94}
For nonzero $\delta$, oxygen vacancies
are present in the CuO layer and the remaining oxygen can
disorder onto sites perpendicular to the CuO chains, thereby
reducing the orthorhombic distortion. \cite{tal89}  However,
occupation of these off-chain sites is inhibited by repulsive
interactions with oxygen located at on-chain sites.  The formation
of twin boundaries, which are abrupt boundaries between crystal
domains with CuO chain orientations that differ by $90^{\circ}$,
are the energetic compromise between decreasing the orthorhombicity
and maintaining the CuO chain ordering.

The overdoped YBa$_2$Cu$_3$O$_{6.993}$ crystal studied by
Hosseini {\it et al.\/} was prepared by
annealing an as-grown crystal at $350^{\circ}$C for
50 days in flowing high purity oxygen. \cite{hos99}  Detwinning
was accomplished by incrementally applying uniaxial stress in the
$\widehat{ab}$ plane with the sample temperature fixed at $250^{\circ}$C
in flowing high purity oxygen.  Progress was monitored by
viewing the sample through a microscope objective with a polarizer
oriented at $45^{\circ}$ with respect to the crystalline axes.  Domains
with differing CuO chain orientations were then visible and the
elimination of twin boundaries could be verified.  The complete
detwinning of a 1~mm$^2$ crystal can typically be accomplished
within one day.

\begin{figure}[tb]
\centerline{\psfig{file=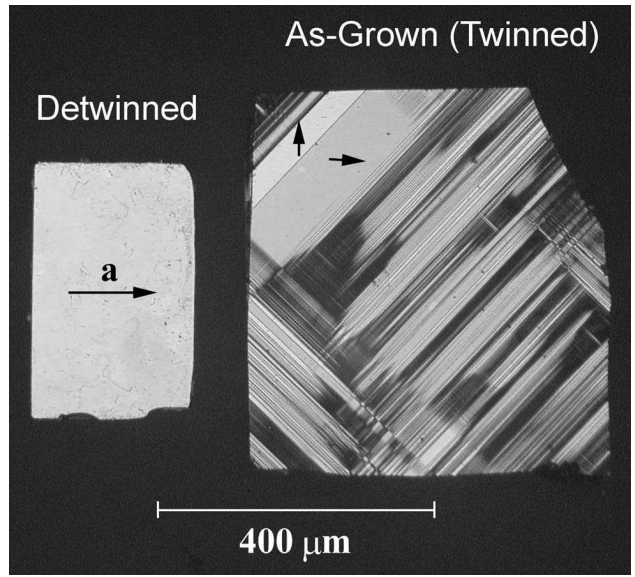}}
\vspace{0.5cm}
\caption{Photograph of as-grown (twinned) and detwinned single crystals
of YBCO as viewed through an optical microscope under polarized light.
In the as-grown sample, two different CuO chain orientations, visible
as a difference in grayscale, are separated by twin boundaries oriented at
$\pm 45^{\circ}$ to the horizontal.  (Arrows highlight the two CuO chain
orientations.)  Darkened areas are regions of highly
concentrated twin boundaries.  In the detwinned sample, twin boundaries
are no longer visible and the $\hat{a}$ axis is parallel to the horizontal
throughout the crystal.  (Photograph courtesy of R. Harris. \cite{HAR01})}
\label{fig:twinpicture}
\end{figure}

An example of an as-grown twinned YBa$_2$Cu$_3$O$_{7-\delta}$ sample and a
detwinned sample of YBa$_2$Cu$_3$O$_{6.993}$ are shown in
Fig.~\ref{fig:twinpicture}.  This picture was obtained by shining polarized
white light onto the crystals at $45^{\circ}$ with respect to the
crystalline axes and then observing through a polarizer oriented
parallel to the incident light.  In the as-grown sample, the two different
CuO chain orientations are visible as a difference in grayscale.  The lines
oriented at $\pm 45^{\circ}$ which separate these regions are 
the twin boundaries.  Darkened areas denote regions of
highly concentrated twin boundaries.
In the detwinned sample, there is a single orientation for all the CuO
chains.  Here, the $\hat{a}$ axis
is parallel to the horizontal since this is the direction
along which the uniaxial stress was applied.  Note the absence of
any visible twin boundaries in the detwinned crystal.

\section{Extended Linear Defect Scattering}
\label{sec:extended}
Since line defect scattering has the appealing property that
it does not sample the full quasiparticle density of states,
we consider a model in which $d$-wave quasiparticles scatter
from extended linear defects.  In particular, we imagine that
the line defects are arranged in a manner reflective of the
twin boundary structure exhibited in YBCO prior to detwinning.
That is, we consider domains of line defects within which the
lines are parallel, oriented at either $+45^{\circ}$ or
$-45^{\circ}$ to the crystal axes, and separated by distances
on the order of microns.  Our picture is that in the process
of detwinning, neighboring twin boundaries are made to
annihilate in order to eliminate regions in which the CuO chain
orientation is disfavored by the applied uniaxial stress.  Although
this procedure is effective in removing the twin boundaries,
it can leave behind defects along the lines at which the twin
boundaries annihilate.
While there may be multiple ways in which this can come
to be, one example of a detwinning scenario through which line
defects could be left behind is the following.

The twin boundaries in as-grown YBCO are $\pm 45^{\circ}$ lines
which separate regions with horizontal CuO chains from regions
with vertical CuO chains.  Geometry dictates that the distance
between oxygen sites on opposite sides of such a line is smaller
than the normal oxygen-oxygen separation in the bulk.  As a result,
it is energetically advantageous for oxygen vacancies, which must be
present in the CuO layer of doped YBCO, to be concentrated along
twin boundary lines.  Upon detwinning, a uniaxial stress is applied
to favor one CuO chain orientation and force neighboring twin
boundaries to approach each other and annihilate.  As the twin
boundaries move, the vacancies move with them until neighboring
boundaries annihilate, leaving behind lines of oxygen vacancies.
The resulting potential, felt by quasiparticles in the CuO$_{2}$
plane, is that of a collection of extended linear defects arranged
in a pattern that reflects the original twinning structure.
Note that this is just one example of a process that could yield
line defects.  As we shall see, however, our results will suggest
that something similar to this is going on.

\begin{figure}[tb]
\centerline{\psfig{file=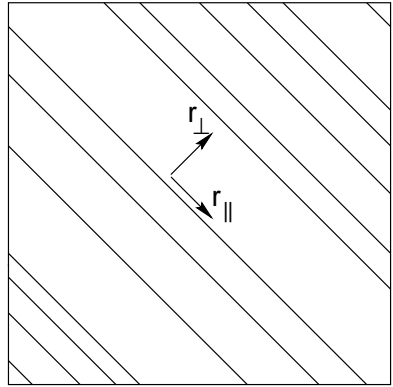}}
\vspace{0.5cm}
\caption{Schematic depiction of a single domain of extended
linear defects.  The line defects are aligned at $-45^{\circ}$ to
the horizontal and spaced randomly in the $\hat{\bf r}_{\perp}$
direction with a 1d density $n_{T}$.}
\label{fig:rtbdomain}
\end{figure}

Consider a single domain in which all line defects are aligned at $-45^{\circ}$
to the horizontal axis.  This situation is depicted schematically in
Fig.\ \ref{fig:rtbdomain}.  We denote the directions parallel and
perpendicular to the defect lines as $\hat{\bf r}_{\parallel}$ and
$\hat{\bf r}_{\perp}$ respectively.  Now consider the effect of
scattering off of such extended linear defects in 2d.  Since all the lines are
normal to the $r_{\perp}$ direction, they act as potential sources,
$V(r_{\perp})$, that depend only on the $r_{\perp}$
coordinate.  Fourier transforming, it is clear that such scattering events
conserve the parallel component of momentum, $p_{\parallel}$, and are therefore
one-dimensional in nature.  Furthermore, since the distribution of line
defects is random in the $r_{\perp}$ direction, this problem corresponds to an
effective one-dimensional ``impurity'' problem (along $\hat{\bf r}_{\perp}$)
in which $p_{\parallel}$ is conserved but not relevant to the scattering.
Performing a 1d disorder average, properties of the resulting system
can be calculated in terms of the 1d line defect density, $n_{T}$, and the 1d
momentum-space line defect scattering potential, $V(p_{\perp}-k_{\perp})$.

\begin{figure}[tb]
\centerline{\psfig{file=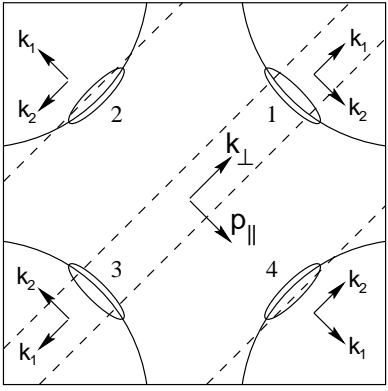}}
\vspace{0.5cm}
\caption{Schematic depiction of the Brillouin zone of a $d$-wave
superconductor in the presence of line defect scattering.  The elliptical
regions, labeled 1 through 4, denote (at a very exaggerated scale)
four pockets of thermally generated nodal quasiparticles.  In
addition to the global momentum directions, $\hat{\bf p}_{\parallel}$
and $\hat{\bf k}_{\perp}$, defined parallel and perpendicular to the
defect lines, we have also defined local momentum directions,
$\hat{\bf k}_{1}$ and $\hat{\bf k}_{2}$, about each of the four nodes.
The dashed lines denote scattering paths that conserve the parallel
component of momentum.}
\label{fig:bz}
\end{figure}

Now consider the effect of this sort of one-dimensional scattering on
the physics of a $d$-wave superconductor.  We model a generic $d$-wave
superconductor as a system with the Brillouin zone of a 2d square
lattice and an order parameter of $d_{x^{2}-y^{2}}$ symmetry that
vanishes at each of four nodal points on the Fermi surface.  In the
vicinity of each of these nodal points, the electronic dispersion,
$\epsilon_{k}$, varies linearly across the Fermi surface and the
order parameter, $\Delta_{k}$, varies linearly along the Fermi surface.
As a result, near each of the gap nodes, the Bogoliubov quasiparticle
excitation spectrum takes the anisotropic Dirac form
$E_{k}=\sqrt{\epsilon_{k}^{2}+\Delta_{k}^{2}}=
\sqrt{v_{f}^{2} k_{1}^{2}+v_{2}^{2} k_{2}^{2}}$
where the degree of anisotropy is expressed by the ratio of the
Fermi velocity, $v_{f}$, to the gap velocity (slope), $v_{2}$,
and the momenta, $k_{1}$ and $k_{2}$, are defined locally at
each node such that in all cases, $\epsilon_{k}=v_{f}k_{1}$
and $\Delta_{k}=v_{2}k_{2}$.
Hence, excitations are free at the
nodal points and at temperatures much less than the gap maximum,
quasiparticles are thermally generated within small regions
about the gap nodes.
The situation is depicted in Fig.\ \ref{fig:bz} where the
elliptical regions (labeled 1 through 4) denote four pockets of
thermally generated quasiparticles.  In the case of point defect
(impurity) scattering, quasiparticles could be scattered
either within the nodal pocket from which they originated
(intra-node) or from one node to another (inter-node). \cite{dur00}
This is so because point defect scattering events do not
conserve either component of momentum.  However, due to the
1d nature of extended linear defect scattering, scattering
events conserve the parallel component of momentum, $p_{\parallel}$.
Therefore, in the presence of line defect scattering, quasiparticles
can only be scattered along lines parallel to the $k_{\perp}$
axis in momentum space.  Four such (dashed) lines are shown in
Fig.\ \ref{fig:bz}.  From the figure it is clear that the condition
of $p_{\parallel}$ conservation means that we have two different
kinds of nodes.  For the odd nodes (1 and 3), quasiparticles can
be scattered either intra-node ($1 \rightarrow 1$ or $3 \rightarrow 3$)
or to their opposite-node ($1 \rightarrow 3$ or $3 \rightarrow 1$).
Forbidden by $p_{\parallel}$ conservation is the adjacent-node scattering
(i.e. $1 \rightarrow 2$) that would be allowed for the point defect case.
But for the even nodes (2 and 4), quasiparticles can only be
scattered intra-node ($2 \rightarrow 2$ or $4 \rightarrow 4$).  Here
both adjacent-node and opposite-node scattering are forbidden by
$p_{\parallel}$ conservation.  Due to the clear difference between
odd nodes and even nodes in the presence of line defect scattering,
it will be important in the calculations that
follow to always treat the odd node and even node cases separately.
Furthermore, we should note that the designation of odd or even to a
particular node is a direct result of our choice to consider a domain
in which the line defects are aligned at $-45^{\circ}$ to the horizontal.
To consider the other type of domain, where the line defects are aligned at
$+45^{\circ}$ to the horizontal, we need only swap odd for even
in all designations.  With this model of thermally excited pockets
of quasiparticles scattered by parallel randomly-spaced defect lines
in mind, we proceed to calculate the resulting self-energy and
microwave conductivity.

\section{Self-Energy Calculation}
\label{sec:selfenergy}
For a superconductor at finite temperature, our calculations will
employ Matsubara Green's functions expressed in the $2 \times 2$
matrix Nambu formalism.  The bare Green's function takes the form
\begin{equation}
\tilde{\cal{G}}_{0}({\bf k},i\omega) =
\frac{1}{(i\omega)^{2} - E_{k}^{2}}
\left( \begin{array}{cc} i\omega + \epsilon_{k} & \Delta_{k} \\
\Delta_{k} & i\omega - \epsilon_{k} \end{array} \right)
\label{eq:baregreenfunc}
\end{equation}
where the tilde denotes a Nambu-space matrix and
$i\omega = i(2n+1)\pi k_{B}T$ is a fermionic Matsubara frequency.
In the presence of scattering, the Green's function is dressed
via Dyson's equation
\begin{equation}
\tilde{\cal{G}}({\bf k},i\omega)^{-1} =
\tilde{\cal{G}}_{0}({\bf k},i\omega)^{-1} -
\tilde{\Sigma}({\bf k},i\omega)
\label{eq:dyson}
\end{equation}
where
\begin{equation}
\tilde{\Sigma}({\bf k},i\omega) =
\tilde{\Sigma}(k_{\perp},k_{\parallel},i\omega) = 
\left( \begin{array}{cc} \Sigma_{11} & \Sigma_{12} \\
\Sigma_{21} & \Sigma_{22} \end{array} \right)
\label{eq:selfenergydef}
\end{equation}
is the Matsubara self-energy.

\begin{figure}[b]
\centerline{\psfig{file=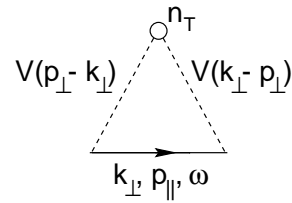}}
\vspace{0.5cm}
\caption{Self-energy within Born approximation.}
\label{fig:bornselfenergy}
\end{figure}

In the Born approximation, this self-energy matrix can be calculated
by evaluating the diagram in Fig.\ \ref{fig:bornselfenergy}
to obtain
\begin{equation}
\tilde{\Sigma}(p_{\perp},p_{\parallel},i\omega) =
n_{T} \sum_{k_{\perp}} \left| V(p_{\perp}-k_{\perp}) \right|^{2}
\tilde{\tau}_{3} \tilde{\cal{G}}(k_{\perp},p_{\parallel},i\omega)
\tilde{\tau}_{3}
\label{eq:Sigmatildegeneral}
\end{equation}
where $V(p_{\perp}-k_{\perp})$ is the line defect scattering potential,
$n_{T}$ is the line defect density, and
the $\tilde{\tau}_{i}$ are Pauli matrices in Nambu-space.
Due to the 1d nature of the line defect scattering, $p_{\parallel}$
is conserved and we have integrated only over $k_{\perp}$.  Since
quasiparticles reside only in the vicinity of the four gap
nodes, each momentum can be expressed by a node index, $j$,
and a local momentum, $(k_{1},k_{2})$, defined about node $j$.
Furthermore, it is useful to scale out the anisotropy of
the excitation spectrum by defining local scaled momenta,
$k^{\prime}_{1} \equiv v_{f} k_{1}$ and
$k^{\prime}_{2} \equiv v_{2} k_{2}$, such that at each node
$\epsilon_{k}=k^{\prime}_{1}$, $\Delta_{k}=k^{\prime}_{2}$, and
$E_{k}=\sqrt{k^{\prime 2}_{1}+k^{\prime 2}_{2}}$.
For convenience, we drop the primes in all that follows and take
all locally defined momenta to be scaled momenta.  Making
use of these locally defined variables, we can replace
\begin{equation}
\sum_{k_{\perp}} \rightarrow \sum_{j}^{\prime}
\int_{-\infty}^{\infty} \frac{dk}{2 \pi v}
\label{eq:sumreplacement}
\end{equation}
where the sum over node index $j$ is restricted to include only
nodes that are crossed by the $k_{\perp}$ integration path,
$k$ is the local scaled momentum component parallel to $k_{\perp}$
($k_{1}$ or $k_{2}$), and $v$ is the corresponding velocity
($v_{f}$ or $v_{2}$).  Similarly, we can replace the initial
momentum, ($p_{\perp}$,$p_{\parallel}$), with an initial node
index and an initial local scaled momentum, ($p_{1}$,$p_{2}$).
As is clear from the dashed integration paths shown in
Fig.\ \ref{fig:bz}, the details of these replacements
depend on the type of node about which the quasiparticles
reside prior to scattering.
For quasiparticles initially near odd nodes (1 or 3), the node
index sum yields both an intra-node scattering term and an
opposite-node scattering term, in both of which $k_{1}$ is
the local integration variable and $p_{2}$ is the local
conserved variable.  In contrast, for quasiparticles initially
near even nodes (2 or 4), the node index sum yields only an
intra-node scattering term, for which $k_{2}$ is the
local integration variable and $p_{1}$ is the local conserved
variable.  It is therefore clear that once we change to local
coordinates, we obtain from Eq.~(\ref{eq:Sigmatildegeneral})
two self-energy expressions, one for odd nodes and one for even
nodes.
For odd nodes:
\begin{eqnarray}
\tilde{\Sigma}^{o}(p_{2},i\omega) = \frac{n_{T}}{2 v_{f}}
\tilde{\tau}_{3} \int \frac{dk_{1}}{\pi}
&& \left[ V_{1}^{2} \tilde{\cal{G}}^{o}(k_{1},p_{2},i\omega)
\right. \nonumber \\
&& \left. + V_{3}^{2} \tilde{\cal{G}}^{o}(k_{1},-p_{2},i\omega)
\right] \tilde{\tau}_{3}
\label{eq:Sigmaodd1}
\end{eqnarray}
whereas for even nodes:
\begin{equation}
\tilde{\Sigma}^{e}(p_{1},i\omega) = \frac{n_{T}}{2 v_{2}}
\tilde{\tau}_{3} \int \frac{dk_{2}}{\pi} \left[ V_{1}^{2}
\tilde{\cal{G}}^{e}(p_{1},k_{2},i\omega) \right]
\tilde{\tau}_{3}
\label{eq:Sigmaeven1}
\end{equation}
where the superscripts `o' and `e' denote odd and even respectively.
Here we have parameterized the scattering potential using the
notation defined in Ref.\ \onlinecite{dur00} whereby $V_{1}$ defines
the potential for an intra-node scattering event and $V_{3}$
defines the potential for an opposite-node scattering event.
Note that for both odd nodes and even nodes, the self-energy is
a function of frequency as well as the conserved component of the
local scaled momentum, $p_{2}$ for the odd case and $p_{1}$ for
the even case, but never the momentum component along $k_{\perp}$.
This is a tremendous simplification since it means that the
self-energy functions built into the right-hand side of Eqs.~(\ref{eq:Sigmaodd1})
and (\ref{eq:Sigmaeven1}) will not be functions of the integration
variables and can always be treated as constants with respect to the
integrals.  It is this fact that makes these equations tractable.

Since the bare Green's function, Eq.~(\ref{eq:baregreenfunc}), has no
$\tilde{\tau}_{2}$ component, it is clear from the form of
Eqs.~(\ref{eq:dyson}) and (\ref{eq:Sigmatildegeneral}) that it is
self-consistent for the self-energy matrix to lack a
$\tilde{\tau}_{2}$ component as well.  It is therefore convenient
to write the odd-node and even-node self-energy matrices in the form
\begin{equation}
\tilde{\Sigma}^{j} = \Sigma^{j} \tilde{\openone} +
\Sigma_{1}^{j} \tilde{\tau}_{3} +
\Sigma_{2}^{j} \tilde{\tau}_{1}
\label{eq:selfenergyredef}
\end{equation}
where $j=\{o,e\}$.
It then follows from Eq.~(\ref{eq:dyson}) that the corresponding dressed
Green's functions can be expressed as
\begin{equation}
\tilde{\cal{G}}^{j}(q_{1},q_{2},i\omega) =
\frac{(i\omega - \Sigma^{j}) \tilde{\openone} +
(q_{1} + \Sigma_{1}^{j}) \tilde{\tau}_{3} +
(q_{2} + \Sigma_{2}^{j}) \tilde{\tau}_{1}}{(i\omega - \Sigma^{j})^{2} -
(q_{1} + \Sigma_{1}^{j})^{2} - (q_{2} + \Sigma_{2}^{j})^{2}}
\label{eq:dressedgreenfunc}
\end{equation}
where $(q_{1},q_{2})$ is the scaled momentum about a particular node
and the self-energy components are functions of $q_{2}$ for odd nodes
and $q_{1}$ for even nodes.
This expression for the Green's functions in terms of the self-energies
together with Eqs.\ (\ref{eq:Sigmaodd1}) and (\ref{eq:Sigmaeven1}),
giving the self-energies in terms of the Green's functions, comprise
matrix equations which can be solved self-consistently for the odd-node
and even-node self-energies.

Let us consider the odd case first.  Shifting the integration
variable $k_{1}$ by $\Sigma^{o}_{1}$, we see that the
$\tilde{\tau}_{3}$ term integrates to zero and therefore
$\Sigma^{o}_{1}=0$.  Then continuing $i\omega\rightarrow\omega+i\delta$
and making the ansatz (which will soon be shown valid)
that $\Sigma^{o}$ is an even function of $p_{2}$ while
$\Sigma^{o}_{2}$ is an odd function of $p_{2}$, we obtain
\begin{mathletters}
\label{eq:Sigmaoboth}
\begin{equation}
\Sigma^{o} = \int \frac{dk_{1}}{\pi}
\frac{\alpha^{o} (\omega - \Sigma^{o})}
{(\omega-\Sigma^{o})^{2} - (p_{2} + \Sigma^{o}_{2})^{2} - k_{1}^{2}}
\label{eq:Sigmao}
\end{equation}
\begin{equation}
\Sigma^{o}_{2} = \int \frac{dk_{1}}{\pi}
\frac{-\alpha^{o}_{1} (p_{2} + \Sigma^{o}_{2})}
{(\omega-\Sigma^{o})^{2} - (p_{2} + \Sigma^{o}_{2})^{2} - k_{1}^{2}}
\label{eq:Sigmao2}
\end{equation}
\begin{equation}
\alpha^{o} \equiv \frac{n_{T} (V_{1}^{2} + V_{3}^{2})}{2 v_{f}}
\;\;\;\;\;\;
\alpha^{o}_{1} \equiv \frac{n_{T} (V_{1}^{2} - V_{3}^{2})}{2 v_{f}}
\label{eq:alphaodef}
\end{equation}
\end{mathletters}
where $\Sigma^{o}=\Sigma^{o}(p_{2},\omega)$
and $\Sigma^{o}_{2}=\Sigma^{o}_{2}(p_{2},\omega)$ are now
retarded functions.

For even nodes, we follow an analogous procedure and obtain
results of the same form.  This time, shifting the integration variable
$k_{2}$ by $\Sigma^{e}_{2}$, it is the $\tilde{\tau}_{1}$ term that
integrates to zero such that $\Sigma^{e}_{2}=0$.  Continuing
$i\omega\rightarrow\omega+i\delta$ to obtain retarded functions
$\Sigma^{e}=\Sigma^{e}(p_{1},\omega)$ and
$\Sigma^{e}_{1}=\Sigma^{e}_{1}(p_{1},\omega)$, we find
\begin{mathletters}
\label{eq:Sigmaeboth}
\begin{equation}
\Sigma^{e} = \int \frac{dk_{2}}{\pi}
\frac{\alpha^{e} (\omega - \Sigma^{e})}
{(\omega-\Sigma^{e})^{2} - (p_{1} + \Sigma^{e}_{1})^{2} - k_{2}^{2}}
\label{eq:Sigmae}
\end{equation}
\begin{equation}
\Sigma^{e}_{1} = \int \frac{dk_{2}}{\pi}
\frac{\alpha^{e}_{1} (p_{1} + \Sigma^{e}_{1})}
{(\omega-\Sigma^{e})^{2} - (p_{1} + \Sigma^{e}_{1})^{2} - k_{2}^{2}}
\label{eq:Sigmae1}
\end{equation}
\begin{equation}
\alpha^{e} \equiv \alpha^{e}_{1} \equiv
\frac{n_{T} V_{1}^{2}}{2 v_{2}}
\label{eq:alphaedef}
\end{equation}
\end{mathletters}
where we have defined $\alpha^{e}_{1}$ in analogy with the odd
node case.

In light of the similarities between the odd and even results,
we can treat both cases on the same footing by defining for
$j=\{\mbox{odd},\mbox{even}\}$: $k=\{k_{1},k_{2}\}$,
$p=\{p_{2},p_{1}\}$, $\Sigma=\{\Sigma^{o},\Sigma^{e}\}$,
$\Sigma_{1}=\{\Sigma^{o}_{2},\Sigma^{e}_{1}\}$,
$\alpha=\{\alpha^{o},\alpha^{e}\}$,
$\alpha_{1}=\{\alpha^{o}_{1},\alpha^{e}_{1}\}$,
and $\eta=\{-1,+1\}$.  Hence, keeping in mind that our new variables
have different meanings for odd and even nodes, we can write
\begin{mathletters}
\label{eq:Sigmajboth}
\begin{equation}
\Sigma = \int \frac{dk}{\pi}
\frac{\alpha (\omega - \Sigma)}
{(\omega-\Sigma)^{2} - (p + \Sigma_{1})^{2} - k^{2}}
\label{eq:Sigmaj}
\end{equation}
\begin{equation}
\Sigma_{1} = \int \frac{dk}{\pi}
\frac{\eta \alpha_{1} (p + \Sigma_{1})}
{(\omega-\Sigma)^{2} - (p + \Sigma_{1})^{2} - k^{2}}
= \eta \frac{\alpha_{1}}{\alpha}
\frac{p + \Sigma_{1}}{\omega - \Sigma} \Sigma
\label{eq:Sigmaj1}
\end{equation}
\end{mathletters}
where the second equality in Eq.~(\ref{eq:Sigmaj1}) was achieved by
making use of Eq.~(\ref{eq:Sigmaj}).  Solving Eq.~(\ref{eq:Sigmaj1})
for $\Sigma_{1}$ in terms of $\Sigma$, plugging the result
into Eq.~(\ref{eq:Sigmaj}), and evaluating the integral we find
\begin{equation}
\Sigma = \frac{-i \alpha}{ \left[ 1 - \left(
\frac{p}{\omega - \gamma \Sigma} \right)^{2} \right]^{1/2}}
\;\;\;\;\;\;
\Sigma_{1} = \frac{(\gamma - 1) p}{\omega - \gamma \Sigma} \Sigma
\label{eq:SigmaSigma1exact}
\end{equation}
where we have defined
$\gamma \equiv 1 + \eta \alpha_{1} / \alpha$.  This equation
will be solved numerically in Sec.~\ref{ssec:numerical} to
obtain an exact result for the self-energy.

Fortunately, in the experimentally relevant ``thermal regime'',
$\Omega,1/\tau \ll T \ll \Delta_{0}$, we can extract an
approximate analytic result.  As will be shown in the microwave
conductivity calculation discussed in the following section,
the energies of interest are $|\omega|$ on the order of $T$
and what we have called $\alpha$ is on the order of $1/\tau$.
Thus, in the thermal regime, we are justified in taking the limit
$|\omega| \gg \alpha$ such that we can set
$\omega-\gamma\Sigma \approx \omega$.  As a result,
defining real and imaginary parts via
$\Sigma = \Lambda - i\Gamma$ and
$\Sigma_{1} = \Lambda_{1} - i\Gamma_{1}$, we find the approximate
expressions
\begin{mathletters}
\label{eq:SigmaSigma1approx}
\begin{equation}
\Gamma \approx \alpha \frac{\theta (|\omega|-|p|)}
{\sqrt{1 - \left( \frac{p}{\omega} \right)^{2}}}
\;\;\;\;\;\;
\Lambda \approx -\alpha\, \mbox{sgn} (\omega)
\frac{\theta (|p|-|\omega|)}
{\sqrt{\left( \frac{p}{\omega} \right)^{2} - 1}}
\label{eq:GammaLambdaapprox}
\end{equation}
\begin{equation}
\Gamma_{1} \approx \eta \alpha_{1} \frac{p}{\omega}
\frac{\theta (|\omega|-|p|)}
{\sqrt{1 - \left( \frac{p}{\omega} \right)^{2}}}
\;\;\;\;\;\;
\Lambda_{1} \approx -\eta \alpha_{1} \frac{p}{|\omega|}
\frac{\theta (|p|-|\omega|)}
{\sqrt{\left( \frac{p}{\omega} \right)^{2} - 1}}
\label{eq:Gamma1Lambda1approx}
\end{equation}
\end{mathletters}
where the sharp cutoffs of the theta-functions are clearly
artifacts of our approximation that are smoothed away in
the exact solution.  In the following section, we will
make use of these thermal-regime self-energy functions
to obtain an analytic expression for the microwave conductivity
in the experimentally relevant parameter regime.

\section{Microwave Conductivity Calculation}
\label{sec:microcond}
\begin{figure}[tb]
\centerline{\psfig{file=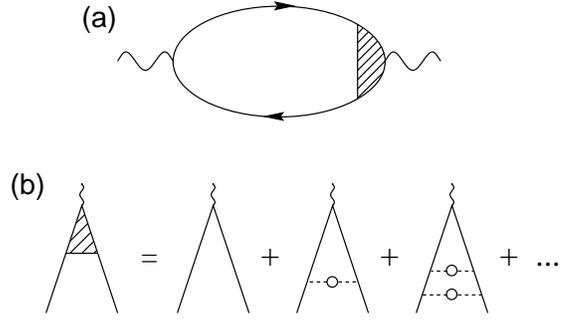}}
\vspace{0.5cm}
\caption{Polarization bubble and dressed vertex within
Born approximation.}
\label{fig:bornbubble}
\end{figure}

The electrical conductivity tensor can be calculated by means
of the Kubo formula
\begin{equation}
\tensor{\sigma}(\Omega,T) =
-\frac{\mbox{Im}\, \tensor{\Pi}_{ret}(\Omega)}{\Omega}
\label{eq:kubo}
\end{equation}
where $\tensor{\Pi}_{ret}(\Omega)
=\tensor{\Pi}(i\Omega\rightarrow\Omega+i\delta)$
and $\tensor{\Pi}(i\Omega)$ is the Matsubara polarization function
(or current-current correlation function).  Evaluating the
diagram in Fig.~\ref{fig:bornbubble}(a) we find
\begin{eqnarray}
\tensor{\Pi}(i\Omega)
&=& \frac{1}{\beta} \sum_{i\omega} \sum_{p}
e^{2} v_{f}^{2} \hat{\bf v}_{fp} \nonumber \\
&\times& \mbox{Tr} \left[ \tilde{\cal{G}}({\bf p},i\omega)
\tilde{\cal{G}}({\bf p},i\omega+i\Omega)
\tilde{\bf \Gamma}({\bf p},i\omega,i\Omega) \right]
\label{eq:bub1}
\end{eqnarray}
where $\hat{\bf v}_{fp}$ points in the direction of the Fermi velocity
and $\tilde{\bf \Gamma}$ is the dressed
vertex function.  Including vertex corrections within the
Born approximation, $\tilde{\bf \Gamma}$ is calculated by
evaluating the ladder diagrams in Fig.~\ref{fig:bornbubble}(b).
Doing so, we obtain
\begin{eqnarray}
\tilde{\bf \Gamma} && (p_{\perp},p_{\parallel},i\omega,i\Omega)
= \hat{\bf v}_{fp} \tilde{\openone} + \sum_{k_{\perp}}
n_{T} \left| V(p_{\perp} - k_{\perp}) \right|^{2} \nonumber \\
&& \times \tilde{\tau}_{3}
\tilde{\cal{G}}(k_{\perp},p_{\parallel},i\omega+i\Omega)
\tilde{\bf \Gamma}(k_{\perp},p_{\parallel},i\omega,i\Omega)
\tilde{\cal{G}}(k_{\perp},p_{\parallel},i\omega)
\tilde{\tau}_{3} \nonumber \\ &&
\label{eq:vertex1}
\end{eqnarray}
where $n_{T}$ is the line defect density and $V(p_{\perp}-k_{\perp})$
is the line defect scattering potential.  Once again, due to the 1d nature
of our scattering, the momentum component parallel to the defect lines,
$p_{\parallel}$, is conserved and we only integrate over the
perpendicular component, $k_{\perp}$.  Replacing momentum
integrals by sums over node index and integrals over local
scaled momenta, the polarization tensor takes the form
\begin{eqnarray}
\tensor{\Pi} && (i\Omega) = e^{2} v_{f}^{2}
\sum_{j=1}^{4} \hat{\bf v}_{f}^{j} \hat{\bf v}_{f}^{j}
\frac{1}{\beta} \sum_{i\omega}
\int \frac{d^{2}p}{(2\pi)^{2}v_{f}v_{2}} \nonumber \\
&& \times \mbox{Tr} \left[ \tilde{\cal{G}}^{j}({\bf p},i\omega)
\tilde{\cal{G}}^{j}({\bf p},i\omega+i\Omega)
\left( \tilde{\openone}+\tilde{\Lambda}^{j}({\bf p},i\omega,i\Omega)
\right) \right] \nonumber \\ &&
\label{eq:bub2}
\end{eqnarray}
where we have defined a vertex correction function, $\tilde{\Lambda}^{j}$,
such that
\begin{equation}
\tilde{\bf \Gamma}^{j}({\bf p},i\omega,i\Omega) = \hat{\bf v}_{f}^{j}
\left( \tilde{\openone} +
\tilde{\Lambda}^{j}({\bf p},i\omega,i\Omega) \right) .
\label{eq:GammaLambdadef}
\end{equation}
Note that $\tilde{\Lambda}^{j}$ depends
only on the conserved component
of the scaled local momentum (i.e.\
$\tilde{\Lambda}^{o}=\tilde{\Lambda}^{o}(p_{2},i\omega,i\Omega)$
while $\tilde{\Lambda}^{e}=\tilde{\Lambda}^{e}(p_{1},i\omega,i\Omega)$).
Since quasiparticles in odd nodes (1 or 3) can be scattered
either intra-node or to the opposite-node while quasiparticles
in even nodes (2 or 4) can only be scattered intra-node, we
must treat the odd and even cases separately as we proceed to
calculate the vertex correction.

For odd nodes, Eqs.~(\ref{eq:vertex1}) and
(\ref{eq:GammaLambdadef}) yield
\begin{eqnarray}
\tilde{\Lambda}^{o}(p_{2}) &=& \frac{n_{T} V_{1}^{2}}{2 v_{f}}
\int \frac{dk_{1}}{\pi} \tilde{\tau}_{3}
\tilde{\cal{G}}^{o}(k_{1},p_{2},i\omega+i\Omega) \nonumber \\
&\times& \left( \tilde{\openone} + \tilde{\Lambda}^{o}(p_{2}) \right)
\tilde{\cal{G}}^{o}(k_{1},p_{2},i\omega) \tilde{\tau}_{3} \nonumber \\
&-& \frac{n_{T} V_{3}^{2}}{2 v_{f}}
\int \frac{dk_{1}}{\pi} \tilde{\tau}_{3}
\tilde{\cal{G}}^{o}(k_{1},-p_{2},i\omega+i\Omega) \nonumber \\
&\times& \left( \tilde{\openone} + \tilde{\Lambda}^{o}(-p_{2}) \right)
\tilde{\cal{G}}^{o}(k_{1},-p_{2},i\omega) \tilde{\tau}_{3}
\label{eq:oddLambda1}
\end{eqnarray}
where we have suppressed the frequency dependence of
$\tilde{\Lambda}^{o}$ for simplicity.  Taking the Nambu-space
trace, noting that
$\tilde{\tau}_{3} \tilde{\cal{G}}^{o}(p_{2}) \tilde{\tau}_{3}
= \tilde{\cal{G}}^{o}(-p_{2})$,
and using the cyclic properties of the trace, we find
\begin{eqnarray}
\mbox{Tr} \left[ \tilde{\Lambda}^{o}(p_{2}) \right] =
\mbox{Tr} \bigg[ && \frac{n_{T}}{2v_{f}} \tilde{I}^{o}(p_{2})
\Big( (V_{1}^{2}-V_{3}^{2}) \tilde{\openone} \nonumber \\
&& + V_{1}^{2} \tilde{\Lambda}^{o}(p_{2}) -
V_{3}^{2} \tilde{\tau}_{3} \tilde{\Lambda}^{o}(-p_{2})
\tilde{\tau}_{3} \Big) \bigg]
\label{eq:troddLambda1}
\end{eqnarray}
where we have defined
\begin{eqnarray}
\tilde{I}^{o}(p_{2}) &\equiv& \int \frac{dk_{1}}{\pi}
\tilde{\cal{G}}^{o}(k_{1},p_{2},i\omega)
\tilde{\cal{G}}^{o}(k_{1},p_{2},i\omega+i\Omega) \nonumber \\
&\equiv& I^{o}(p_{2}) \tilde{\openone}
+ I_{1}^{o}(p_{2}) \tilde{\tau}_{1}
\label{eq:Iodef}
\end{eqnarray}
with $I^{o}(p_{2})$ an even function and $I^{o}_{1}(p_{2})$ an
odd function.  Note that the fact that $\tilde{I}^{o}(p_{2})$
can be written in this form is merely assumed at this point
but will be demonstrated later.
Manipulating further and making the additional assumption
(which will soon be shown valid) that
$\mbox{Tr}[\tilde{\Lambda}^{o}(p_{2})]$ is an even function and
$\mbox{Tr}[\tilde{\tau}_{1}\tilde{\Lambda}^{o}(p_{2})]$ is an
odd function, we can write
\begin{equation}
\mbox{Tr} \left[ \tilde{\Lambda}^{o} \right]
= \alpha_{1}^{o} I^{o} \left( 2 +
\mbox{Tr} \left[ \tilde{\Lambda}^{o} \right] \right)
+ \alpha_{1}^{o} I^{o}_{1}
\,\mbox{Tr} \left[ \tilde{\tau}_{1} \tilde{\Lambda}^{o} \right]
\label{eq:traceeq1}
\end{equation}
where $\alpha_{1}^{o}=n_{T}(V_{1}^{2}-V_{3}^{2})/2v_{f}$,
as defined in Eq.~(\ref{eq:alphaodef}), and 
we have now suppressed the $p_{2}$ dependences.  Going back to
Eq.~(\ref{eq:oddLambda1}) but this time multiplying by
$\tilde{\tau}_{1}$ before taking the trace, we can proceed
along analogous lines to show that
\begin{equation}
\mbox{Tr} \left[ \tilde{\tau}_{1} \tilde{\Lambda}^{o} \right]
= \alpha_{1}^{o} J^{o}
\,\mbox{Tr} \left[ \tilde{\tau}_{1} \tilde{\Lambda}^{o} \right]
+ \alpha_{1}^{o} J^{o}_{1}
\left( 2 + \mbox{Tr} \left[ \tilde{\Lambda}^{o} \right] \right)
\label{eq:traceeq2}
\end{equation}
where we have now defined
\begin{eqnarray}
\tilde{J}^{o}(p_{2}) &\equiv& -\int \frac{dk_{1}}{\pi}
\tilde{\cal{G}}^{o}(k_{1},p_{2},i\omega) \tilde{\tau}_{1}
\tilde{\cal{G}}^{o}(k_{1},p_{2},i\omega+i\Omega) \nonumber \\
&\equiv& J^{o}_{1}(p_{2}) \tilde{\openone}
+ J^{o}(p_{2}) \tilde{\tau}_{1}
\label{eq:Jodef}
\end{eqnarray}
with $J^{o}(p_{2})$ an even function and $J^{o}_{1}(p_{2})$ an
odd function.  Solving Eqs.~(\ref{eq:traceeq1}) and (\ref{eq:traceeq2})
simultaneously, we find that
\begin{equation}
\mbox{Tr} \left[ \tilde{\Lambda}^{o} \right] = \frac{2
\alpha_{1}^{o} \left( I^{o} +
\frac{\alpha_{1}^{o} I^{o}_{1} J^{o}_{1}}{1 - \alpha_{1}^{o} J^{o}}
\right)}{1 -
\alpha_{1}^{o} \left( I^{o} +
\frac{\alpha_{1}^{o} I^{o}_{1} J^{o}_{1}}{1 - \alpha_{1}^{o} J^{o}}
\right)}
\label{eq:troddLambda2}
\end{equation}
and
\begin{equation}
\mbox{Tr} \left[ \tilde{\tau}_{1} \tilde{\Lambda}^{o} \right] =
\frac{\frac{2 \alpha_{1}^{o} J^{o}_{1}}{1 - \alpha_{1}^{o} J^{o}}}
{1 - \alpha_{1}^{o} \left( I^{o} +
\frac{\alpha_{1}^{o} I^{o}_{1} J^{o}_{1}}{1 - \alpha_{1}^{o} J^{o}}
\right)}
\label{eq:trtau1Lambda}
\end{equation}
from which it is clear, given the defined parity of $I^{o}$, $I^{o}_{1}$,
$J^{o}$, and $J^{o}_{1}$, that as a function of $p_{2}$,
$\mbox{Tr}[\tilde{\Lambda}^{o}]$ is odd and
$\mbox{Tr}[\tilde{\tau}_{1}\tilde{\Lambda}^{o}]$ is even, in agreement
our prior assumptions.  Finally, using Eqs.~(\ref{eq:Iodef}) and
(\ref{eq:traceeq1}) and the fact that the Pauli matrices are traceless,
we can write
\begin{equation}
\mbox{Tr} \left[ \tilde{\Lambda}^{o}(p_{2}) \right] = \alpha_{1}^{o}
\,\mbox{Tr} \left[ \tilde{I}^{o}(p_{2}) \left( \tilde{\openone} +
\tilde{\Lambda}^{o}(p_{2}) \right) \right]
\label{eq:troddLambda3}
\end{equation}
which will prove a useful expression in what follows.

For even nodes, the vertex correction function takes the form
\begin{eqnarray}
\tilde{\Lambda}^{e}(p_{1}) &=& \frac{n_{T} V_{1}^{2}}{2 v_{2}}
\int \frac{dk_{2}}{\pi} \tilde{\tau}_{3}
\tilde{\cal{G}}^{e}(p_{1},k_{2},i\omega+i\Omega) \nonumber \\
&\times& \left( \tilde{\openone} + \tilde{\Lambda}^{e}(p_{1}) \right)
\tilde{\cal{G}}^{e}(p_{1},k_{2},i\omega) \tilde{\tau}_{3} . \nonumber \\
\label{eq:evenLambda1}
\end{eqnarray}
Note that this even-node expression is of the same form as the
corresponding odd-node expression, Eq.~(\ref{eq:oddLambda1}), but is somewhat
simpler since we need only consider intra-node scattering.  Taking the
trace and cyclically permuting within the trace, we find
\begin{equation}
\mbox{Tr} \left[ \tilde{\Lambda}^{e}(p_{1}) \right] = \alpha_{1}^{e}
\,\mbox{Tr} \left[ \tilde{I}^{e}(p_{1}) \left( \tilde{\openone} +
\tilde{\Lambda}^{e}(p_{1}) \right) \right]
\label{eq:trevenLambda1}
\end{equation}
where $\alpha_{1}^{e}=n_{T}V_{1}^{2}/2v_{f}$,
as defined in Eq.~(\ref{eq:alphaedef}), and we have defined
\begin{eqnarray}
\tilde{I}^{e}(p_{1}) &\equiv& \int \frac{dk_{2}}{\pi}
\tilde{\cal{G}}^{e}(p_{1},k_{2},i\omega)
\tilde{\cal{G}}^{e}(p_{1},k_{2},i\omega+i\Omega) \nonumber \\
&\equiv& I^{e}(p_{1}) \tilde{\openone}
+ I_{1}^{e}(p_{1}) \tilde{\tau}_{3} .
\label{eq:Iedef}
\end{eqnarray}
Making use of the second equality in Eq.~(\ref{eq:Iedef})
and further defining
\begin{eqnarray}
\tilde{J}^{e}(p_{1}) &\equiv& \int \frac{dk_{2}}{\pi}
\tilde{\cal{G}}^{e}(p_{1},k_{2},i\omega) \tilde{\tau}_{3}
\tilde{\cal{G}}^{e}(p_{1},k_{2},i\omega+i\Omega) \nonumber \\
&\equiv& J^{e}_{1}(p_{1}) \tilde{\openone}
+ J^{e}(p_{1}) \tilde{\tau}_{3}
\label{eq:Jedef}
\end{eqnarray}
we obtain two coupled equations for
$\mbox{Tr}[\tilde{\Lambda}^{e}]$ and
$\mbox{Tr}[\tilde{\tau}_{3}\tilde{\Lambda}^{e}]$,
\begin{equation}
\mbox{Tr} \left[ \tilde{\Lambda}^{e} \right]
= \alpha_{1}^{e} I^{e} \left( 2 +
\mbox{Tr} \left[ \tilde{\Lambda}^{e} \right] \right)
+ \alpha_{1}^{e} I^{e}_{1}
\,\mbox{Tr} \left[ \tilde{\tau}_{3} \tilde{\Lambda}^{e} \right]
\label{eq:eventraceeq1}
\end{equation}
\begin{equation}
\mbox{Tr} \left[ \tilde{\tau}_{3} \tilde{\Lambda}^{e} \right]
= \alpha_{1}^{e} J^{e}
\,\mbox{Tr} \left[ \tilde{\tau}_{3} \tilde{\Lambda}^{e} \right]
+ \alpha_{1}^{e} J^{e}_{1}
\left( 2 + \mbox{Tr} \left[ \tilde{\Lambda}^{e} \right] \right) .
\label{eq:eventraceeq2}
\end{equation}
Solving simultaneously yields
\begin{equation}
\mbox{Tr} \left[ \tilde{\Lambda}^{e} \right] = \frac{2
\alpha_{1}^{e} \left( I^{e} +
\frac{\alpha_{1}^{e} I^{e}_{1} J^{e}_{1}}{1 - \alpha_{1}^{e} J^{e}}
\right)}{1 -
\alpha_{1}^{e} \left( I^{e} +
\frac{\alpha_{1}^{e} I^{e}_{1} J^{e}_{1}}{1 - \alpha_{1}^{e} J^{e}}
\right)}
\label{eq:trevenLambda2}
\end{equation}
which has precisely the same form as in the odd case.

Due to this similarity in form, we can unite the odd and even cases
by using the superscript $j=\{o,e\}$ and defining a generalized integration
variable $k=\{k_{1},k_{2}\}$ and a generalized conserved
variable $p=\{p_{2},p_{1}\}$.  Doing so,
we can define
\begin{eqnarray}
K^{j}(p,i\omega,i\Omega) &\equiv& \frac{1}{2\alpha_{1}^{j}}
\mbox{Tr} \left[ \tilde{\Lambda}^{j}(p,i\omega,i\Omega) \right] \nonumber \\
&=& \frac{I^{j} +
\frac{\alpha_{1}^{j} I^{j}_{1} J^{j}_{1}}{1 - \alpha_{1}^{j} J^{j}}}
{1 - \alpha_{1}^{j} \left( I^{j} +
\frac{\alpha_{1}^{j} I^{j}_{1} J^{j}_{1}}{1 - \alpha_{1}^{j} J^{j}}
\right)}
\label{eq:Kdef}
\end{eqnarray}
and note via Eqs.~(\ref{eq:troddLambda3}) and (\ref{eq:trevenLambda1}) that
\begin{equation}
\mbox{Tr} \left[ \tilde{I}^{j}(p,i\omega,i\Omega) \left( \tilde{\openone}
+ \tilde{\Lambda}^{j}(p,i\omega,i\Omega) \right) \right]
= 2 K^{j}(p,i\omega,i\Omega) .
\label{eq:trLambdaj}
\end{equation}
Then going back to our expression for the polarization tensor,
Eq.~(\ref{eq:bub2}), and noting that
\begin{eqnarray}
\hat{\bf v}_{f}^{1} \hat{\bf v}_{f}^{1}
+ \hat{\bf v}_{f}^{3} \hat{\bf v}_{f}^{3}
&=& \tensor{\openone} + \tensor{\tau}_{1}
\nonumber \\
\hat{\bf v}_{f}^{2} \hat{\bf v}_{f}^{2}
+ \hat{\bf v}_{f}^{4} \hat{\bf v}_{f}^{4}
&=& \tensor{\openone} - \tensor{\tau}_{1}
\label{eq:vvtensor}
\end{eqnarray}
we can write the tensor as the sum of an odd-node term and
an even-node term
\begin{equation}
\tensor{\Pi}(i\Omega) =
\Pi^{o}(i\Omega) \left[ \tensor{\openone} + \tensor{\tau}_{1} \right] +
\Pi^{e}(i\Omega) \left[ \tensor{\openone} - \tensor{\tau}_{1} \right]
\label{eq:buboddeven}
\end{equation}
where, for $j=\{o,e\}$,
\begin{eqnarray}
\Pi^{j}(i\Omega) && = \frac{e^{2}}{4\pi} \frac{v_{f}}{v_{2}}
\int \!dp\, \frac{1}{\beta} \sum_{i\omega}
\mbox{Tr}\bigg[ \int\!\frac{dk}{\pi} \tilde{\cal{G}}^{j}(k,p,i\omega)
\nonumber \\
&& \;\;\;\;\;\;\;\;\;\;\;\;\;\;\;
\times \tilde{\cal{G}}^{j}(p,k,i\omega+i\Omega)
\left( \tilde{\openone}+\tilde{\Lambda}^{j}(p,i\omega,i\Omega)
\right) \bigg] \nonumber \\
&& = \frac{e^{2}}{2\pi} \frac{v_{f}}{v_{2}}
\int \!dp\, \frac{1}{\beta} \sum_{i\omega} K^{j}(p,i\omega,i\Omega).
\label{eq:bubj}
\end{eqnarray}
At this point, we would like to evaluate the Matsubara sum of
$K^{j}(p,i\omega,i\Omega)$, analytically continue the Matsubara
frequencies, and obtain retarded functions.  To do so we should
note that the internal and external frequencies,
$i\omega$ and $i\Omega$, enter $K^{j}(p,i\omega,i\Omega)$
only through functions of ``Matsubara-couplets'' of the form
$A(i\omega)B(i\omega+i\Omega)$, where both A and B have
the analytic structure of a Matsubara Green's function.
A procedure for evaluating Matsubara sums of functions of
this type has been developed in Appendix B of
Ref.~\onlinecite{dur00}.  Applying this procedure to the
case at hand, we find that the imaginary part of the
retarded polarization function takes the form
\begin{eqnarray}
\mbox{Im}\,\Pi^{j}_{ret}(\Omega) = && \frac{e^{2}}{2\pi}
\frac{v_{f}}{v_{2}} \int dp
\int \frac{d\omega}{2\pi}
\left( n_{F}(\omega+\Omega)-n_{F}(\omega) \right) \nonumber \\
&& \times \mbox{Re} \left[ K^{j}_{B}(p,\omega,\Omega) -
K^{j}_{A}(p,\omega,\Omega) \right]
\label{eq:imretbubj}
\end{eqnarray}
where $n_{F}$ is the Fermi function and
\begin{eqnarray}
K^{j}_{A}(p,\omega,\Omega) &=& \lim_{i\Omega\rightarrow\Omega+i\delta}
K^{j}(p,\omega+i\delta,i\Omega) \nonumber \\
K^{j}_{B}(p,\omega,\Omega) &=& \lim_{i\Omega\rightarrow\Omega+i\delta}
K^{j}(p,\omega-i\delta,i\Omega)
\label{eq:Kab}
\end{eqnarray}
are what we will refer to as the A-form and B-form of
$K^{j}(p,i\omega,i\Omega)$.  Plugging into the Kubo
formula, Eq.~(\ref{eq:kubo}), this yields the conductivity tensor
\begin{equation}
\tensor{\sigma}(\Omega,T) =
\sigma^{o}(\Omega,T) \left[ \tensor{\openone} + \tensor{\tau}_{1} \right] +
\sigma^{e}(\Omega,T) \left[ \tensor{\openone} - \tensor{\tau}_{1} \right]
\label{eq:sigmaoddeven}
\end{equation}
where
\begin{eqnarray}
\sigma^{j} && (\Omega,T) = \frac{e^{2}}{4\pi^{2}}
\frac{v_{f}}{v_{2}} \int_{-\infty}^{\infty} \!d\omega
\left( \frac{n_{F}(\omega)-n_{F}(\omega+\Omega)}{\Omega} \right) \nonumber \\
&& \times \int_{-\infty}^{\infty} \!dp\
\mbox{Re} \left[ K^{j}_{B}(p,\omega,\Omega) -
K^{j}_{A}(p,\omega,\Omega) \right] .
\label{eq:sigmajKK}
\end{eqnarray}

The next step is to express $K^{j}_{A}$ and $K^{j}_{B}$ in terms of
the self-energy calculated in Sec.~\ref{sec:selfenergy}.
To do so we must first evaluate our expressions for the
Matsubara-formalism functions,
$I^{j}$, $I_{1}^{j}$, $J^{j}$, and $J_{1}^{j}$,
and then analytically continue to obtain the corresponding
A-forms and B-forms.  Dressed via the self-energy
functions, the Matsubara Green's functions, evaluated at
frequencies $i\omega$ and $i\omega+i\Omega$, take the form
\begin{mathletters}
\label{eq:oddevenGreenfunc}
\begin{equation}
\tilde{\cal{G}}^{o}(k,p,i\omega)
= \frac{f_{1}^{o}\tilde{\openone} + k\tilde{\tau}_{3}
+ g_{1}^{o}\tilde{\tau}_{1}}{\left. f_{1}^{o} \right.^{2}
- k^{2} - \left. g_{1}^{o} \right.^{2}}
\end{equation}
\begin{equation}
\tilde{\cal{G}}^{o}(k,p,i\omega+i\Omega)
= \frac{f_{2}^{o}\tilde{\openone} + k\tilde{\tau}_{3}
+ g_{2}^{o}\tilde{\tau}_{1}}{\left. f_{2}^{o} \right.^{2}
- k^{2} - \left. g_{2}^{o} \right.^{2}}
\end{equation}
for odd nodes and
\begin{equation}
\tilde{\cal{G}}^{e}(p,k,i\omega)
= \frac{f_{1}^{e}\tilde{\openone} + g_{1}^{e}\tilde{\tau}_{3}
+ k\tilde{\tau}_{1}}{\left. f_{1}^{e} \right.^{2}
- \left. g_{1}^{e} \right.^{2} - k^{2}}
\end{equation}
\begin{equation}
\tilde{\cal{G}}^{e}(p,k,i\omega+i\Omega)
= \frac{f_{2}^{e}\tilde{\openone} + g_{2}^{e}\tilde{\tau}_{3}
+ k\tilde{\tau}_{1}}{\left. f_{2}^{e} \right.^{2}
- \left. g_{2}^{e} \right.^{2} - k^{2}}
\end{equation}
\end{mathletters}
for even nodes, where we have defined
\begin{mathletters}
\label{eq:f1f2g1g2}
\begin{eqnarray}
f_{1}^{j} &\equiv& i\omega - \Sigma^{j}(p,i\omega) \\
f_{2}^{j} &\equiv& i\omega+i\Omega - \Sigma^{j}(p,i\omega+i\Omega) \\
g_{1}^{j} &\equiv& p + \Sigma_{1}^{j}(p,i\omega) \\
g_{2}^{j} &\equiv& p + \Sigma_{1}^{j}(p,i\omega+i\Omega) .
\end{eqnarray}
\end{mathletters}
Plugging the dressed Green's functions into
Eqs.~(\ref{eq:Iodef}), (\ref{eq:Jodef}), (\ref{eq:Iedef}),
and (\ref{eq:Jedef}), we find that
\begin{mathletters}
\label{eq:MatsuII1JJ1}
\begin{equation}
I(p,i\omega,i\Omega) = \int \frac{dk}{\pi}
\frac{f_{1} f_{2} + g_{1} g_{2} + k^{2}}
{(f_{1}^{2} - g_{1}^{2} - k^{2})
(f_{2}^{2} - g_{2}^{2} - k^{2})}
\end{equation}
\begin{eqnarray}
I_{1}(p,i\omega,i\Omega) &=& \eta J_{1}(p,i\omega,i\Omega) \nonumber \\
&=& \int \frac{dk}{\pi}
\frac{f_{1} g_{2} + f_{2} g_{1}}
{(f_{1}^{2} - g_{1}^{2} - k^{2})
(f_{2}^{2} - g_{2}^{2} - k^{2})} \nonumber \\ & &
\end{eqnarray}
\begin{equation}
J(p,i\omega,i\Omega) = \eta \int \frac{dk}{\pi}
\frac{f_{1} f_{2} + g_{1} g_{2} - k^{2}}
{(f_{1}^{2} - g_{1}^{2} - k^{2})
(f_{2}^{2} - g_{2}^{2} - k^{2})}
\end{equation}
\end{mathletters}
where $\eta=\{-1,1\}$ for $j=\{o,e\}$ and we have suppressed
the node-index superscripts for simplicity.
Expanding the integrands via partial fraction decomposition,
noting that the resulting integrals are precisely of the form
that appear in Eq.~(\ref{eq:Sigmajboth}),
and using this to relate the integrals to our self-energy
functions, the above can be evaluated.  Then analytically
continuing as prescribed in (\ref{eq:Kab}),
we obtain both the A-form functions
\begin{mathletters}
\label{eq:AformIJ}
\begin{equation}
I_{A} = \frac{1}{\alpha}
\frac{(f_{2}+f_{1})(\Sigma-\Sigma_{+})
+ \eta \frac{\alpha}{\alpha_{1}} (g_{2}-g_{1})(\Sigma_{1}+\Sigma_{1+})}
{(f_{2}^{2}-f_{1}^{2}) - (g_{2}^{2}-g_{1}^{2})}
\end{equation}
\begin{equation}
I_{1A} = \frac{1}{\alpha}
\frac{(g_{2}\Sigma - g_{1}\Sigma_{+})
+ \eta \frac{\alpha}{\alpha_{1}} (f_{2}\Sigma_{1} - f_{1}\Sigma_{1+})}
{(f_{2}^{2}-f_{1}^{2}) - (g_{2}^{2}-g_{1}^{2})}
\end{equation}
\begin{equation}
J_{A} = \frac{\eta}{\alpha}
\frac{(f_{2}-f_{1})(\Sigma+\Sigma_{+})
+ \eta \frac{\alpha}{\alpha_{1}} (g_{2}+g_{1})(\Sigma_{1}-\Sigma_{1+})}
{(f_{2}^{2}-f_{1}^{2}) - (g_{2}^{2}-g_{1}^{2})}
\end{equation}
\end{mathletters}
and the B-form functions
\begin{mathletters}
\label{eq:BformIJ}
\begin{equation}
I_{B} = \frac{1}{\alpha}
\frac{(f_{2}+f_{1}^{*})(\Sigma^{*}-\Sigma_{+})
+ \eta \frac{\alpha}{\alpha_{1}} (g_{2}-g_{1}^{*})(\Sigma_{1}^{*}+\Sigma_{1+})}
{(f_{2}^{2}-f_{1}^{*2}) - (g_{2}^{2}-g_{1}^{*2})}
\end{equation}
\begin{equation}
I_{1B} = \frac{1}{\alpha}
\frac{(g_{2}\Sigma^{*} - g_{1}^{*}\Sigma_{+})
+ \eta \frac{\alpha}{\alpha_{1}} (f_{2}\Sigma_{1}^{*} - f_{1}^{*}\Sigma_{1+})}
{(f_{2}^{2}-f_{1}^{*2}) - (g_{2}^{2}-g_{1}^{*2})}
\end{equation}
\begin{equation}
J_{B} = \frac{\eta}{\alpha}
\frac{(f_{2}-f_{1}^{*})(\Sigma^{*}+\Sigma_{+})
+ \eta \frac{\alpha}{\alpha_{1}} (g_{2}+g_{1}^{*})(\Sigma_{1}^{*}-\Sigma_{1+})}
{(f_{2}^{2}-f_{1}^{*2}) - (g_{2}^{2}-g_{1}^{*2})}
\end{equation}
\end{mathletters}
where all self-energies are now retarded and
$\Sigma \equiv \Sigma(p,\omega)$,
$\Sigma_{+} \equiv \Sigma(p,\omega+\Omega)$,
$\Sigma_{1} \equiv \Sigma_{1}(p,\omega)$,
$\Sigma_{1+} \equiv \Sigma_{1}(p,\omega+\Omega)$,
$f_{1}=\omega-\Sigma$,
$f_{2}=\omega+\Omega-\Sigma_{+}$,
$g_{1}=p+\Sigma_{1}$, and
$g_{2}=p+\Sigma_{1+}$.
Then, in terms of the A-form and B-form functions above,
\begin{equation}
K_{(A,B)} = \frac{I_{(A,B)}
+ \frac{\eta \alpha_{1} I_{1(A,B)}^{2}}{1 - \alpha_{1} J_{(A,B)}}}
{1 - \alpha_{1} \left( I_{(A,B)}
+ \frac{\eta \alpha_{1} I_{1(A,B)}^{2}}{1 - \alpha_{1} J_{(A,B)}} \right)} .
\label{eq:ABformK}
\end{equation}
Thus, given the self-energy functions, $\Sigma$ and $\Sigma_{1}$,
Eqs.\ (\ref{eq:sigmaoddeven}), (\ref{eq:sigmajKK}), (\ref{eq:AformIJ}),
(\ref{eq:BformIJ}), and (\ref{eq:ABformK})
constitute a well-defined prescription for calculating the
conductivity tensor.  In Sec.~\ref{ssec:numerical}, these equations
will be evaluated numerically, in conjunction with a numerical
solution for the self-energy, in order to obtain an exact
result for the conductivity.

However, in the thermal regime,
$\Omega,1/\tau \ll T \ll \Delta_{0}$,
we are justified in taking the limit
$\Omega,\alpha,\alpha_{1} \ll |\omega|$,
which simplifies our results significantly and allows for
an analytic solution.  Taking this limit and using the
thermal regime self-energy calculated in Sec.~\ref{sec:selfenergy},
we find that the B-form results become
\begin{equation}
I_{B} \approx \frac{\omega}{p} I_{1B}
\approx \eta\left(\frac{\omega}{p}\right)^{2} \! J_{B}
\approx \frac{\frac{1}{\alpha}}{1 + \eta\frac{\alpha_{1}}{\alpha}
\left(\frac{p}{\omega}\right)^{2} - i\frac{\Omega}{2\Gamma}}
\label{eq:IBI1BJB}
\end{equation}
where
\begin{equation}
\Gamma \approx \alpha \frac{\theta (|\omega|-|p|)}
{\sqrt{1 - \left( \frac{p}{\omega} \right)^{2}}}
\label{eq:Gammaapprox}
\end{equation}
as expressed in Eq.~(\ref{eq:SigmaSigma1approx}).
Remarkably, these functions conspire to give $K_{B}$
the surprisingly simple form
\begin{equation}
K_{B} \approx \frac{2 \tau_{tr}}
{1 - i\Omega\tau_{tr}\frac{\alpha}{\Gamma}}
\label{eq:KB}
\end{equation}
where we have suggestively defined an effective transport
scattering rate
\begin{equation}
\frac{1}{2\tau_{tr}} \equiv \alpha - \alpha_{1} .
\label{eq:tautrdef}
\end{equation}
Calculating the A-form functions to the same order as in the
the B-form case, we find that
\begin{equation}
I_{A} \approx I_{1A} \approx J_{A} \approx 0
\;\;\; \Longrightarrow \;\;\; K_{A} \approx 0 .
\label{eq:IAI1AJAKA}
\end{equation}
Therefore, taking the real part of $K_{B}-K_{A}$ and integrating
over scaled momentum we see that
\begin{equation}
\int_{-\infty}^{\infty} \!dp\ \mbox{Re}
\left[ K_{B} - K_{A} \right]
= 4 f(\Omega\tau_{tr}) \frac{\tau_{tr}}
{1 + (\Omega\tau_{tr})^{2}} |\omega|
\label{eq:pintKBKA}
\end{equation}
where we have defined the function
\begin{equation}
f(x) \equiv \sqrt{1+\frac{1}{x^{2}}} \ \mbox{arctanh}
\left( \frac{1}{\sqrt{1+\frac{1}{x^{2}}}} \right)
\label{eq:fx}
\end{equation}
which goes as $1-x^{2}/3$ for small argument and as
$\ln (2x)$ for large argument.
Plugging into Eq.~(\ref{eq:sigmajKK}), noting that
\begin{equation}
\int_{-\infty}^{\infty} \!d\omega\ |\omega|
\left( -\frac{\partial n_{F}}{\partial \omega} \right)
= 2 \ln (2) k_{B} T ,
\label{eq:omegaint}
\end{equation}
and writing the node-index superscripts explicitly, yields
the node-$j$ microwave conductivity
\begin{equation}
\sigma^{j}(\Omega,T) = \frac{e^{2}}{\pi^{2}} \frac{v_{f}}{v_{2}}
2 \ln (2) f(\Omega\tau_{tr}^{j}) \frac{\tau_{tr}^{j}}
{1 + (\Omega\tau_{tr}^{j})^{2}} k_{B} T .
\label{eq:sigmajfinal}
\end{equation}
It is important to note that for odd nodes
\begin{equation}
\frac{1}{2\tau_{tr}^{o}} = \frac{n_{T}(V_{1}^{2} + V_{3}^{2})}{2v_{f}}
- \frac{n_{T}(V_{1}^{2} - V_{3}^{2})}{2v_{f}}
= \frac{n_{T} V_{3}^{2}}{v_{f}}
\label{eq:tautrodd}
\end{equation}
while for even nodes
\begin{equation}
\frac{1}{2\tau_{tr}^{e}} = \frac{n_{T} V_{1}^{2}}{2v_{2}}
- \frac{n_{T} V_{1}^{2}}{2v_{2}} = 0 .
\label{eq:tautreven}
\end{equation}
Hence, for non-zero frequencies
\begin{equation}
\frac{1}{2\tau_{tr}^{e}} = 0 \;\;\; \Longrightarrow \;\;\;
\sigma^{e}(\Omega,T) = 0
\label{eq:sigmaevenzero}
\end{equation}
so the even nodes make no contribution to the
microwave conductivity.  From a physical point of view, these
results are quite logical.  Since electrical current is
proportional to Fermi velocity, it is directed outward along the
Brillouin zone diagonals at each node
(see Refs.~\onlinecite{lee97,dur00}).
Therefore, quasiparticles in the vicinity of odd nodes carry
current normal to the defect lines which is degraded via
line defect scattering.  However, quasiparticles about even nodes
carry current parallel to the defect lines which is unaffected
by their presence.  Since a nonzero microwave conductivity
requires a nonzero scattering rate, only the odd nodes
can contribute.  Hence
\begin{equation}
\tensor{\sigma}(\Omega,T) =
\sigma^{o}(\Omega,T) \left[ \tensor{\openone} + \tensor{\tau}_{1} \right]
\label{eq:sigmasingledomain}
\end{equation}
which clearly contains off-diagonal components and is therefore
spatially anisotropic.  This is the case because we have been
considering only a single domain in which all defect lines are aligned
at $-45^{\circ}$ to the horizontal.  Since such a system
certainly has a preferred direction, an anisotropic
conductivity makes sense.  However, in reality, we expect
a detwinned crystal to have both $-45^{\circ}$ and $+45^{\circ}$
aligned domains.  As discussed in Sec.~\ref{sec:extended},
our results can be adapted to the $+45^{\circ}$
case by swapping all odd designations for even and vice versa.
Doing so simply changes the sign of the $\tensor{\tau}_{1}$
term in Eq.~(\ref{eq:sigmasingledomain}).  Therefore, averaging
over domains with $\pm 45^{\circ}$ defect lines cancels the off-diagonal
components and yields an isotropic conductivity tensor
\begin{equation}
\tensor{\sigma}(\Omega,T) = \frac{e^{2}}{\pi^{2}} \frac{v_{f}}{v_{2}}
2 \ln (2) f(\Omega\tau_{tr}^{o}) \frac{\tau_{tr}^{o}}
{1 + (\Omega\tau_{tr}^{o})^{2}} k_{B} T \tensor{\openone} .
\label{eq:sigmatensorfinal}
\end{equation}
Hence, within the thermal regime, Born scattering from extended
linear defects in a $d$-wave superconductor yields
both a linear temperature dependence and a near-Drude
frequency dependence.

\section{Results}
\label{sec:results}
\subsection{Analytical Results}
\label{ssec:analytical}
\begin{figure}[tb]
\centerline{\psfig{file=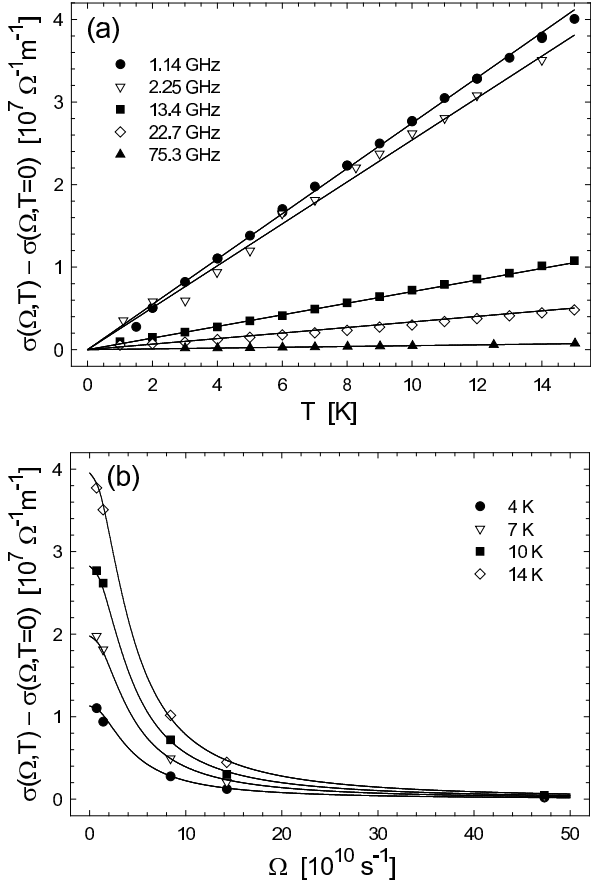}}
\vspace{0.5cm}
\caption{Fits to experiment of thermal regime microwave conductivity
plotted (a) versus temperature for each of five frequencies and
(b) versus frequency for each of four temperatures.  Points denote
the temperature-dependent part of the microwave conductivity
measured in YBa$_{2}$Cu$_{3}$O$_{6.993}$ by Hosseini {\it et al.}
\cite{hos99}  Lines denote a two parameter fit of
Eq.~(\ref{eq:sigmatensorfinal}) obtained with transport lifetime,
$\tau_{tr} = 2.93 \times 10^{-11}$s, and Fermi-liquid renormalization
factor, $\alpha_{fl}^{2} = 0.6$.}
\label{fig:datafit}
\end{figure}

Now let us compare our thermal regime expression for the
microwave conductivity due to extended linear defect scattering,
Eq.~(\ref{eq:sigmatensorfinal}), with the measured microwave
conductivity of detwinned YBa$_{2}$Cu$_{3}$O$_{6.993}$
obtained via experiment by Hosseini {\it et al.} \cite{hos99}
A fit of our calculated result to the temperature-dependent part of
the measured conductivity data is presented in Fig.~\ref{fig:datafit}.
Note that the close agreement seen in the figure was achieved with
only two free parameters: the effective transport lifetime, $\tau_{tr}$,
and an overall scale factor.  The fit yields a lifetime of
$2.93 \times 10^{-11} \mbox{s}$ which corresponds to a mean free path
on the order of microns (quite reasonable for the high-purity
sample in question).  Assuming an anisotropy ratio, $v_{f}/v_{2}=21$,
as measured via sub-Kelvin thermal conductivity by Taillefer
and co-workers \cite{chi00}, we obtain a scale factor of 0.6 by which
our expression must be multiplied to fit the data.  This is in
line with the expected Fermi-liquid correction factor
\cite{mil98,mes98,xu95,dur00}, $\alpha_{fl}^{2} \sim 0.4-0.5$,
that has been obtained from measurements of the superfluid density.
\cite{chi00,zha94,bon96,lee96}  Thus, our thermal regime expression
yields a quantitative fit to the
temperature-dependent part of the measured data.

However, there are several features of the Hosseini data that are not
captured by Eq.~(\ref{eq:sigmatensorfinal}).  First of all, for each
of the frequencies at which data was taken, a temperature-independent
shift was observed in addition to the temperature-dependent part plotted
in Fig.~\ref{fig:datafit}.  Yet in our thermal regime expression,
$\sigma(\Omega,T=0)$ is zero.  Furthermore, though predominantly linear
with temperature, the measured data deviates slightly, but noticeably,
from linearity.  In fact, Hosseini {\it et al.\/} note a gradual
evolution from a concave-down (sub-linear) deviation at low frequencies to a
concave-up (super-linear) deviation at high frequencies.  Our thermal
regime expression is strictly linear with temperature.

From the measured data (see Fig.~4 of Ref.~\onlinecite{hos99}),
it seems clear that the observed concave-down
deviation at low frequencies results from the influence of
inelastic scattering.  For these low frequencies, the conductivity
peak marking the onset of inelastic scattering appears just beyond
the upper bound of our temperature range of interest.
Quite naturally, the onset
of inelastic scattering decreases the conductivity and yields
a concave-down deviation from linearity.  As we are concerned here
with the nature of the low temperature elastic scattering mechanism,
we do not expect to reproduce this feature.  However, the observed
high frequency (concave-up) deviations from linearity, as well as
the temperature-independent shifts, do appear to be features of the
elastic scattering which we seek to understand.

Since Eq.~(\ref{eq:sigmatensorfinal}) is an approximate result,
obtained by taking the limit $1/\tau,\Omega \ll T$
and thereby neglecting sub-dominant terms,
it is possible that sub-dominant features (constant shift
and deviation from linearity for high frequencies) were neglected when
we assumed the thermal limit to obtain our analytical result.  To see
if these features emerge in an exact solution, we present numerical
results, valid beyond the thermal limit, in the following section.

\subsection{Numerical Results}
\label{ssec:numerical}
Recall from Sec.~\ref{sec:selfenergy} that the self-energy in
the presence of line defect scattering is determined by the self-consistent
solution of Eq.~(\ref{eq:SigmaSigma1exact}) as a function of momentum
and energy.  While an analytic expression was obtained in the thermal
limit, this equation must be solved numerically for more general
parameter values.  Doing so (via Newton's method) yields the self-energy
functions, $\Sigma(p,\omega)$ and $\Sigma_{1}(p,\omega)$.  Plugging these
functions into Eqs.~(\ref{eq:sigmaoddeven}), (\ref{eq:sigmajKK}),
(\ref{eq:AformIJ}), (\ref{eq:BformIJ}), and (\ref{eq:ABformK}) and
numerically integrating over momentum and energy, we obtain the
microwave conductivity for a particular temperature and frequency.
Repeating this process for all temperatures and frequencies of interest
yields an exact (numerical) result for the microwave conductivity.

\begin{figure}[tb]
\centerline{\psfig{file=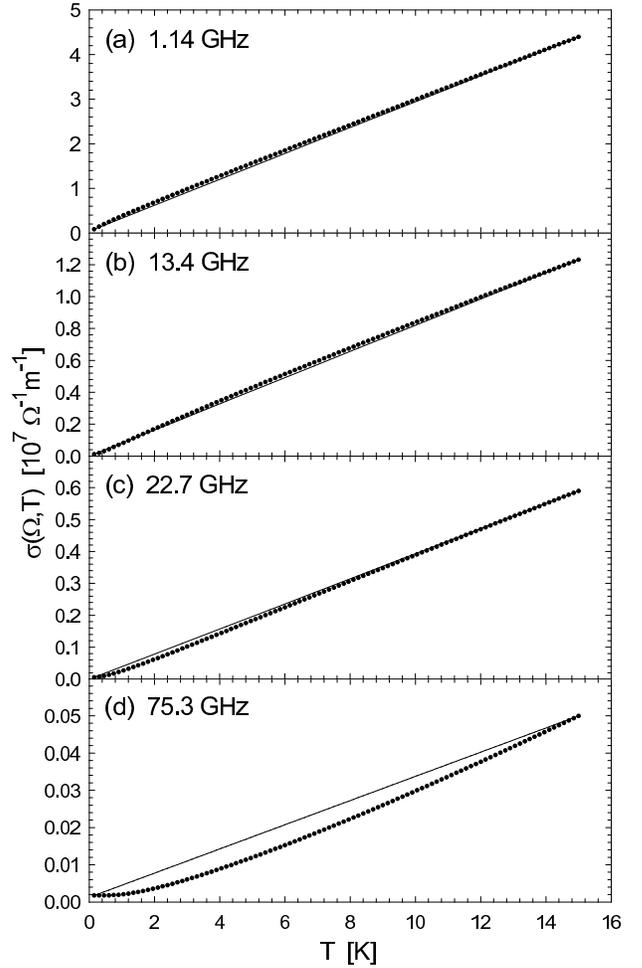}}
\vspace{0.5cm}
\caption{Results of exact numerical calculation of microwave conductivity
plotted as a function of temperature for (a) 1.14 GHz, (b) 13.4 GHz,
(c) 22.7 GHz, and (d) 75.3 GHz.  Straight lines connecting the first
and last data points have been included to highlight any deviation
from linearity.}
\label{fig:numerical}
\end{figure}

Following this procedure, we computed the microwave conductivity as
a function of temperature, from 0 to 15~K, for each of the five
experimentally relevant frequencies.  In all cases, we used values of
the transport lifetime and Fermi-liquid correction factor obtained,
in the previous section, from the fit of our thermal regime expression
to experiment.  The results for 1.14~GHz, 13.4~GHz, 22.7~GHz, and 75.3~GHz are
plotted versus temperature in Fig.~\ref{fig:numerical}.  In each plot,
a straight line has been drawn between the first and last data points
to highlight any deviation from linearity.  These
exact results confirm the conclusions of our thermal regime
calculation, revealing a predominantly linear temperature-dependence
and a near-Drude frequency dependence.  In addition, at the high
frequencies where our thermal regime approximations were least justified,
we obtain, in qualitative agreement with experiment, concave-up
deviations from linearity.
Furthermore, these results do exhibit nonzero, albeit very small,
offsets at zero temperature (most evident for 75.3~GHz).
However, the computed offsets are far smaller than those observed
experimentally.  Perhaps the point defect scattering, which is
certainly present but was ignored herein for simplicity, is
required to reproduce this feature correctly.  A more
complete calculation must consider the effects of both extended
linear defects and point defects in the same system.  This is an
important direction for future study.

\section{Conclusions}
\label{sec:conclusions}
The low temperature behavior of the microwave conductivity measured in
YBa$_{2}$Cu$_{3}$O$_{6.993}$ by Hosseini {\it et al.\/} \cite{hos99} has been
difficult to explain in terms of point defect (impurity) scattering.
The basic problem is that the strong energy dependence of the low
temperature quasiparticle density of states is directly reflected
in a point defect scattering rate which is too energy dependent to
produce a microwave conductivity that agrees with experiment.  This
suggests that an additional low temperature scattering mechanism may
also be important.

We note that extended linear defect scattering is an appealing
candidate.  Unlike point defect scattering, line defect scattering
does not sample the full quasiparticle density of states, so the
line defect scattering rate need not inherit a strong energy dependence.
Furthermore, lines are prevalent in as-grown YBCO in the form of
twin boundaries.  Although these twin boundaries are eliminated
via the application of a uniaxial stress, if remnants of the
twinning structure are left behind, they could take the form
of line defects.

We have calculated (within the Born approximation) the self-energy
and microwave conductivity due to scattering from a single
domain of parallel, randomly spaced line defects, all aligned
at $-45^{\circ}$ to the crystal axes.
We find that the anisotropic nature of line defect
scattering clearly differentiates odd nodes (for which electrical current
is perpendicular to the defect lines) from even nodes (for which electrical
current is parallel to defect lines).  For the former, both intra-node
(forward) scattering and opposite-node (back) scattering
are permitted, whereas for the latter, only intra-node
(forward) scattering is allowed.  By including vertex corrections in
our calculation, we have accounted for the fact that back scattering is
an effective means of degrading a current while forward scattering is not.
Therefore, while odd nodes yield a nonzero transport scattering rate
and microwave conductivity, even nodes make no contribution.
The resulting conductivity tensor is anisotropic, reflective of the
anisotropic nature of the scattering mechanism.  Thus, for a sample
in which all twin boundaries were parallel prior to detwinning, this
anisotropy should be observable.  (In fact, the measurement of the
conductivity tensor in such a single-domain sample would serve as
a good test for the presence of line defects.)  However, since
the samples in question contain multiple domains with line defects
aligned at $\pm 45^{\circ}$, we average over oppositely aligned
domains to recover a conductivity scalar.

In the experimentally relevant limit, $1/\tau, \Omega \ll T \ll \Delta_{0}$,
which we have called the thermal regime, our calculations simplify
and we obtain an analytic expression for the microwave conductivity:
\begin{equation}
\tensor{\sigma}(\Omega,T) = \frac{e^{2}}{\pi^{2}} \frac{v_{f}}{v_{2}}
2 \ln (2) f(\Omega\tau_{tr}^{o}) \frac{\tau_{tr}^{o}}
{1 + (\Omega\tau_{tr}^{o})^{2}} k_{B} T \tensor{\openone} .
\label{eq:sigmatensoragain}
\end{equation}
As this result exhibits a linear temperature dependence and a near-Drude
lineshape (modified logarithmically by the function $f(x)$
defined in Eq.~(\ref{eq:fx})), it captures the most robust qualitative
features of the Hosseini data.  Furthermore, when we fit this expression
to the temperature-dependent part of the measured conductivity
(see Sec.~\ref{ssec:analytical}), we obtain good quantitative agreement
with reasonable values of our two fitting parameters, the transport
lifetime and the Fermi-liquid renormalization factor.  However, we note
that this approximate analytical result fails to reproduce the more
subtle features of the measured data: temperature-independent offsets
and deviations of the temperature dependence from strict linearity
at high frequencies.  (We note that the deviations observed at low
frequencies can be explained by the onset of inelastic scattering.)

In search of these subdominant features, we performed 
an exact numerical calculation, valid beyond the thermal regime.
These numerical results (see Sec.~\ref{ssec:numerical}) do
exhibit deviations from linear temperature dependence at
high frequencies, in qualitative agreement with experiment.
While small temperature-independent offsets are also obtained,
the magnitude of these is far smaller than observed experimentally.
Perhaps the point defect scattering, which we have neglected herein,
must be included to reproduce this feature.

The close agreement of our calculated conductivity with experiment
strongly suggests that extended linear defects are indeed present
in detwinned single crystals of YBa$_{2}$Cu$_{3}$O$_{6.993}$
and make an important contribution to the scattering.
Our picture has been that these line defects are remnants of the twin
boundary structure of the as-grown crystal, left behind after
detwinning.  We described, in Sec.~\ref{sec:extended},
a detwinning scenario whereby the annihilation of twin boundaries
leaves behind lines of oxygen vacancies in the CuO layer.
While this is one example of a process that could yield extended
linear defects, others are certainly possible.  Our results,
however, imply that something similar to this is taking place.

To answer remaining questions and provide a more complete picture
of the low temperature scattering, future calculations should
include the effects of both line defect scattering and point defect
scattering in the same system.  Nevertheless, these initial results
suggest that scattering from extended linear defects has a significant
influence on the observed microwave conductivity in detwinned single
crystals of YBa$_{2}$Cu$_{3}$O$_{6.993}$.

\acknowledgements
The authors would like to thank R. Harris for providing the 
photograph used in Fig.~\ref{fig:twinpicture} as well as the crystal
growth and detwinning information presented in Sec.~\ref{sec:crystal}.
We also gratefully acknowledge useful discussions with
C. Kallin, A. J. Berlinsky, P. J. Hirschfeld, and D. A. Bonn.
This work was supported by NSF DMR-9813764.
We acknowledge the hospitality of the 2000 Boulder Summer School
for Condensed Matter and Materials Physics and of the Institute
for Theoretical Physics at UCSB while this research was ongoing.

\end{document}